\documentclass[runningheads]{llncs}
\usepackage[T1]{fontenc}
\usepackage[colorlinks,linkcolor=green,anchorcolor=green,citecolor=green]{hyperref}
\usepackage{graphicx}
\usepackage{multirow}
\usepackage{diagbox}
\usepackage{threeparttable}
\usepackage{amsfonts,amssymb,amsmath} 
\usepackage[misc]{ifsym}
\usepackage[ruled]{algorithm2e}

\begin{document}
\title{Structure and Smoothness Constrained Dual Networks for MR Bias Field Correction}%
%
\titlerunning{S2DNets: MR Bias Field Correction}
%
\author{Dong Liang\inst{1} \and Xingyu Qiu\inst{1} \and Yuzhen Li\inst{1} \and Wei Wang\inst{1} \and Kuanquan Wang\inst{1} \and Suyu Dong\inst{2} \and Gongning Luo$^{(\textrm{\Letter})}$\inst{1}} 
\authorrunning{D. Liang et al.}

\institute{Harbin Institute of Technology, Harbin, China \\
\and King Abdullah University of Science and Technology, Saudi Arabia
}
%
\maketitle              
\begin{abstract}
MR imaging techniques are of great benefit to disease diagnosis. However, due to the limitation of MR devices, significant intensity inhomogeneity often exists in imaging results, which impedes both qualitative and quantitative medical analysis. Recently, several unsupervised deep learning-based models have been proposed for MR image improvement. However, these models merely concentrate on global appearance learning, and neglect constraints from image structures and smoothness of bias field, leading to distorted corrected results. In this paper, novel structure and smoothness constrained dual networks, named S2DNets, are proposed aiming to self-supervised bias field correction. S2DNets introduce piece-wise structural constraints and smoothness of bias field for network training to effectively remove non-uniform intensity and retain much more structural details. Extensive experiments executed on both clinical and simulated MR datasets show that the proposed model outperforms other conventional and deep learning-based models. In addition to comparison on visual metrics, downstream MR image segmentation tasks are also used to evaluate the impact of the proposed model. The source code is available at:\href{https://github.com/LeongDong/S2DNets}{https://github.com/LeongDong/S2DNets}.

\end{abstract}
\section{Introduction}
Bias field, as an unavoidable challenge in MR image processing tasks, is caused by imperfect MR devices or imaged objects, which introduces intensity inhomogeneity into MR images and compromises subsequent medical analysis methods\cite{r1}\cite{r2}.
\begin{figure}
\includegraphics[width=12cm]{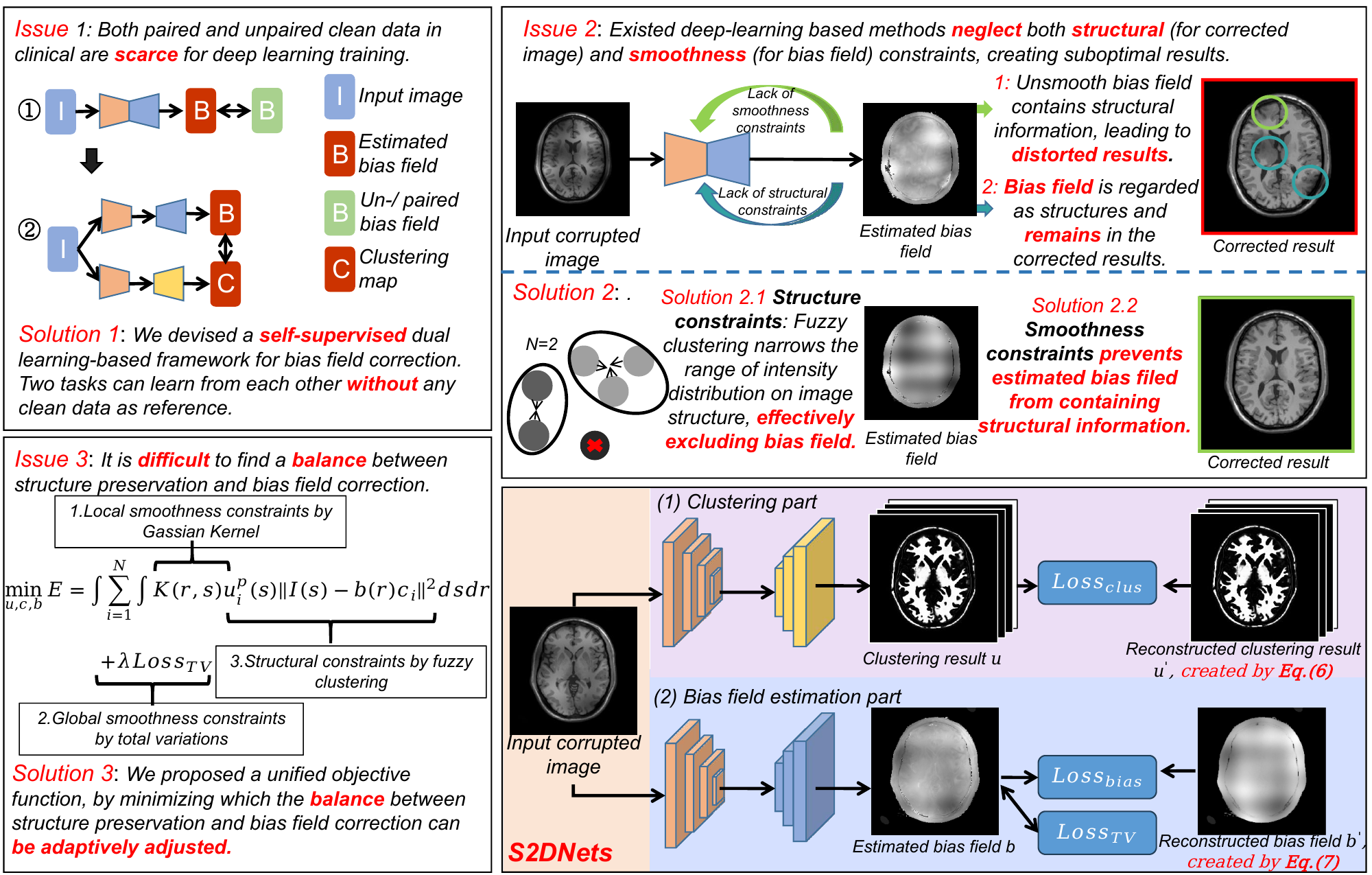}
\centering
\caption{Contributions and framework of proposed model.} \label{fig1}
\end{figure}
In the past decades, various of methods had been proposed to deal with bias field, including histogram-~\cite{r3}\cite{r4}, segmentation-~\cite{r5}-\cite{r8add} and deep learning-based methods~\cite{r9}-\cite{r17}. Smoothness constraints, which describes the fundamental characteristics of bias fields, has been widely adopted in numerous methods~\cite{r3}\cite{r4}\cite{r6}-\cite{r8add}. As bias field varies smoothly in low frequency, Sled et al.~\cite{r3} proposed $N3$ method for bias field correction by iteratively maximizing high frequency within tissues. Tustison et al.~\cite{r4} improved $N3$ to obtain more effective method named $N4$. Mishro et al.~\cite{r5} pointed out that the fuzzy clustering can be used to lower the effect of bias field. Li et al.~\cite{r8} applied low-order polynomial functions to fit smoothly varying bias field. However, it is limited in describing the bias field with complex distribution. Ma et al.~\cite{r29} bypassed explicit modeling of bias field smoothness but instead employed a global hard threshold on image gradients to preserve structural edges for sparse reconstruction. However, due to the lack of smoothness constraints on the bias field, it suffers from an inherent inability to simultaneously resolve the conflict between suppressed weak edges in the restored image and retained spurious gradients arising from bias field artifacts. Some researchers adopted level-set or $L_{0}$ regularization into the bias field estimation framework\cite{r6}~\cite{r8add}. However, these methods rely on manual initialization and cannot be executed automatically. 
 
  Nowadays, deep learning-based methods have achieved remarkable performance in bias field correction tasks. Compared to traditional algorithms, deep learning-based methods offer significant advantages: 1) its end-to-end optimization enables data-driven learning of complex bias field distributions; 2) it provides superior inference efficiency; and 3) it demonstrates enhances generalization capability, maintaining robustness against intensity variation. Both Goldfryd et al.~\cite{r9} and Chen et al.~\cite{r10} adopted GANs model for bias field estimation in a supervised learning scheme. Other GANs-based models like Residual CycleGANs~\cite{r11} and IRNet~\cite{r12}, resorted to learning on unpaired data. However, unpaired clean data is usually absent in clinical practice. Thus, some researchers tried to train network without any clean data as reference. ImplicitNet~\cite{r13}, BECNN~\cite{r14} and DeepN4~\cite{r16} chose to generate artificial bias field as ground truth. However, if the generated result differs from the real data, these models would be degraded significantly. Yang et al.~\cite{r15} adjusted the image intensity as ZeroDCE++~\cite{r18} but it is more suitable for natural image instead of MR image because of different imaging mechanism. Perez-Caballero et al.~\cite{r17} introduced entropy minimization into network. However, it is hard to find a balance between bias field removal and image structure preservation.

To deal with above problems, we propose a novel self-supervised framework for bias field correction, called structure and smoothness constrained dual networks (S2DNets). To this end, we adopted structure constraint to describe features of clean MR image and smoothness constraint to regularize the spatial distribution of bias field, and combine them into a unified objective function. Losses are obtained by solving closed-form solutions of objective function so as to effectively estimate bias field and retain structural details of corrected results as much as possible. S2DNets consist of two sub-networks to predict piece-wise constant structure information by fuzzy clustering and estimate bias field, respectively. Inspired by ~\cite{r19}-\cite{r21}, two sub-networks alternately learn from each other in a self-supervised dual manner. As shown in Fig.~\ref{fig1}, contributions can be summarized as:
\textbf{a)} We propose a novel self-supervised framework called S2DNets for bias field correction. S2DNets introduces a unified objective function by combining structure and smoothness constraints together so as to effectively remove bias field and preserve structural details of corrected results;
\textbf{b)} Extensive experiments conducted on both clinical and simulated dataset demonstrate evident advantages of S2DNets compared with other state-of-the-art methods. 
\section{Method}
Our S2DNets aim to estimate bias field from corrupted MR image. Firstly, the relationship among bias field, corrupted image and clean image is described mathematically, based on which a unified objective function is constructed by incorporating structure and smoothness constraints. Then, closed-form solutions on bias field and clustering are derived from the objective function, which is used to design loss functions for guiding network training. Finally, the details for network architecture is provided.
\subsubsection{2.1 Objective Function for Bias Field Estimation}
In most research~\cite{r1}~\cite{r2}, the relationship between the original MR image and corresponding bias field can be mathematically described in a simple multiplicative form as:
\begin{equation}
I(r)=i(r)b(r)+n(r), \forall r\in\Omega\label{eq1}
\end{equation}
where $\Omega$ is image domain, $r$ denotes the location in the $\Omega$, $I(\cdot)$ is the acquired MR image, $i(\cdot)$ is the corresponding clean MR image without bias field, $b(\cdot)$ represents the bias field, and $n(\cdot)$ is the noise, which could be estimated by quasi-Gaussian functions. As image background often contains irrelevant signals, it may introduce interference for bias field estimation, leading to over-corrected results~\cite{r3}. Thus, in this work, we adopt a multi-threshold OTSU (MOTSU)~\cite{r25} to automatically determine the foreground area as:
\begin{equation}
mask(r)=\left\{
\begin{aligned}
1 & , & if\ I(r)>min(MOTSU(I,M)) \\
0 & , & otherwise
\end{aligned}
\right.\label{eq2}
\end{equation}
where $MOTSU(\cdot)$ outputs computed thresholds, $I$ is acquired MR image, $M$ is the number of threshold, default by 3. $min(\cdot)$ is the function return the minimum value. For that bias field varies smoothly across the whole imaged object, the closer two points are, the more likely they have the same bias field value. Thus, the probability that two points $r$ and $s$ have the same bias field value is approximated by a masked Gaussian kernel $K(r,s)$ as:
\begin{equation}
K(r,s)=\left\{
\begin{aligned}
mask(r)\cdot mask(s)\cdot\frac{1}{\sigma\sqrt{2\pi}}e^{-\frac{(r-s)^{2}}{2\sigma^{2}}} & , & if\ |r-s|<d,\ r,s\in\Omega\\
0&  , & otherwise
\end{aligned}
\right.\label{eq3}
\end{equation}
where $K(r,s)$ is re-normalized as $K(r,s)=K(r,s)/\int K(r,s)ds$, and $d$ is predefined distance, $\sigma$ controls the shape of probability distribution. For a clean MR image, intensity distribution is assumed to be piece-wise constant~\cite{r10}. Thus, pixels with similar intensity can be classified into same cluster by clustering method, and the intensity of each pixel can be conversely approximated by corresponding cluster center. Inspired by fuzzy clustering method\cite{r5}, an objective function is proposed based on Eq.\ref{eq1} and Eq.\ref{eq3} as:
\begin{equation}
\min\limits_{u,c,b} E=\int\sum_{i=1}^{N}\int K(r,s)u_{i}^{p}(s)||I(s)-b(r)c_{i}||^{2}dsdr, \mathrm{s.t.} \sum_{i=1}^{N_{c}}u_{i}(s)=1\label{eq4}
\end{equation}
where $r,s\in\Omega$, $N$ is the number of cluster centers with different gray distributions.
$K(r,s)$ describes the local smoothness of bias field. $u_{i}(s)$ represents the probability that the $s$th pixel is classified into the $i$th cluster, and $c_{i}$ is $i$th cluster center. $p$ is fuzziness factor. $u_{i}^{p}(s)||I(s)-b(r)c_{i}||^{2}$ restricts the corrected result to satisfying the structural constraints that clean MR image is piece-wise constant. For a certain pixel, the sum of probabilities belonging to each cluster is equal to $1$.
\subsubsection{2.2 Closed-form Solutions and Loss Functions}
 Objective function is minimized to estimate the appropriate bias field so as to ensure the smoothness of bias field and preserve structural details of corrected image. Closed-form solutions are acquired by calculating the first derivative equaling zero on variables $u_{i}(r)$, $c_{i}$ and $b(r)$, respectively, as:
\begin{equation}
c_{i}=\frac{\int\int K(r,s)b(r)I(s)u_{i}^{p}(s)dsdr}{\int\int K(r,s)b^{2}(r)u_{i}^{p}(s)dsdr}\label{center}
\end{equation}
\begin{equation}
u_{i}^{'}(r)=\frac{1}{\sum_{j=1}^{N}(\frac{||I(r)-c_{i}\int K(r,s)b(s)ds||^{2}}{||I(r)-c_{j}\int K(r,s)b(s)ds||^{2}})^{\frac{1}{p-1}}}\label{eq5}
\end{equation}
\begin{equation}
b^{'}(r)=\frac{\sum_{i=1}^{N}\int K(r,s)c_{i}I(s)u_{i}^{p}(s)ds}{\sum_{i=1}^{N}\int K(r,s)c_{i}^{2}u_{i}^{p}(s)ds}\label{eq6}
\end{equation}
where $u_{i}^{'}(r)$ and $b^{'}(r)$ are reconstructed probability and bias field, respectively.
For training clustering part, the probability map reconstruction loss function $Loss_{clus}$ is calculated between predicted probability map $u$ and reconstructed probability map $u^{'}$ (see Eq.\ref{eq5}.) as:
\begin{equation}
Loss_{clus}=\sum_{i=1}^{N}\int_{\Omega}||u_{i}^{'}(r)-u_{i}(r)||^{2}dr
\end{equation}
For training bias field estimation part, the bias field reconstruction loss function $Loss_{bias}$ is calculated between predicted bias field $b$ and reconstructed bias field $b^{'}$ (see Eq.\ref{eq6}.) as:
\begin{equation}
Loss_{bias}=\int_{\Omega}||b^{'}(r)-b(r)||^{2}dr
\end{equation}
As $K(r,s)$ only focus on the local smoothness, we propose total variation loss to further introduce the global smoothness constraints of bias field, as:
\begin{equation}
Loss_{TV}=\int_{\Omega}[\partial^{2}_{x}b(r)+\partial^{2}_{y}b(r)]dr
\end{equation}
where $\partial_{x}(\cdot)$ and $\partial_{y}(\cdot)$ are first-order partial derivative along x-axis and y-axis, respectively. The total loss for training bias estimation network is:$Loss_{bias}+\lambda Loss_{TV}$, $\lambda$ is weight for controlling global smoothness of bias field.
\subsubsection{2.3 Network Architecture}
S2DNets adopt two encoder/decoder pairs to estimate probability map \textbf{u} and bias field \textbf{b}, respectively. As shown in Fig.~\ref{fig1}, the proposed model is composed of two parts: a clustering part for predicting the probability of every pixel belonging to each cluster, and a bias estimation part for calculating the bias field, which are all based on the same structure as basic U-Net in~\cite{r22}. The difference is that the clustering part adopts the Softmax function to predict the probability map, and the bias estimation part applies the Sigmoid function to restrict the range of bias field in the last output layer.
\section{Experiment}
\subsubsection{3.1 Datasets}
Experiments are conducted on the clinical dataset~\textbf{HCP}\cite{r23} and simulated dataset~\textbf{BrainWeb}\cite{r24}. In HCP dataset, 30000 T1w slices with size $260\times 311$ were randomly split into 15000/15000 for training/test set. The BrainWeb dataset contains T1 and T2 slice with size $217\times 181$. In BrainWeb, bias field is simulated with strength randomly re-scaling to [0.3,1.7] by Legendre polynomials as~\cite{r10} where 2000/1810 slices were created for training/test set. Center-cropping and zero-padding are used to set the slice size into $256\times 256$. Random flipping and rotation are applied for data augmentation.

\subsubsection{3.2 Implementation Details}
Our proposed network is trained by an Adam optimizer for 1500 iterations with batch size $4$ and initial learning rate is set to 0.001 decayed by 0.999 every iteration. To balance the weight between local and global smoothness constraints, $\lambda=\frac{Loss_{bias}}{Loss_{tv}}$ is adaptively adjusted during training process. The network is based on Pytorch by NVIDIA 2080Ti.\par
Our model is compared with two most widely applied models (i.e., N4~\cite{r4} and MICO~\cite{r8}); four unsupervised models: Residual CycleGAN (abbrev. RCGAN)~\cite{r11} and IR-Net~\cite{r12}, Implicit~\cite{r13} and Yang~\cite{r15}). For training different models, training sets are further randomly splitted into 10000/5000 in HCP and 1810/190 in BrainWeb for corrupted input data and unpaired clean data. Implicit, Yang~\cite{r15} and our S2DNets are trained without any clean data as reference. RCGAN and IRNet are trained with additional unpaired clean data. All codes are executed according to their corresponding literature.  
\begin{table}[]
\caption{Quantitative fidelity comparisons (mean/standard deviation) with Wilcoxon-test between proposed and other methods on both \textbf{HCP} and \textbf{BrainWeb} dataset.} \label{tab1}
\begin{threeparttable}
\begin{tabular}{lcccccc}\\
\hline\hline
\multicolumn{1}{l}{\multirow{2}{*}{Methods}} & \multicolumn{2}{c}{HCP}  & \multicolumn{2}{c}{BrainWeb T1} & \multicolumn{2}{c}{BrainWeb T2} \\
\cline{2-7}
\multicolumn{1}{c}{}                         & PSNR$\uparrow$       & SSIM$\uparrow$        & PSNR$\uparrow$           & SSIM$\uparrow$           & PSNR$\uparrow$           & SSIM$\uparrow$           \\
\cline{1-7}
Original                                     & 19.88$/$8.65$^{\ddag}$ & 0.897$/$0.084$^{\ddag}$ & 23.07$/$3.49$^{\ddag}$     & 0.920$/$0.038$^{\ddag}$    & 20.50$/$3.32$^{\ddag}$     & 0.895$/$0.042$^{\ddag}$    \\
MICO                                         & 19.60$/$8.08$^{\ddag}$ & 0.898$/$0.092$^{\ddag}$ & 19.42$/$3.33$^{\ddag}$     & 0.829$/$0.077$^{\ddag}$    & 16.53$/$3.11$^{\ddag}$     & 0.744$/$0.089$^{\ddag}$    \\
N4                                           & 22.22$/$8.27$^{\ddag}$ & 0.932$/$0.068$^{\ddag}$ & 24.45$/$4.03$^{\ddag}$     & 0.939$/$0.037$^{\ddag}$    & 21.10$/$3.92$^{\ddag}$     & 0.909$/$0.041$^{\ddag}$    \\
Implicit                                     & 22.49$/$5.26$^{\ddag}$ & 0.893$/$0.067$^{\ddag}$ & 15.94$/$3.42$^{\ddag}$     & 0.637$/$0.103$^{\ddag}$    & 16.87$/$1.19$^{\ddag}$     & 0.732$/$0.045$^{\ddag}$    \\
Yang et al.                                  & 23.54$/$6.60$^{\ddag}$ & 0.914$/$0.054$^{\ddag}$ & 25.08$/$3.15$^{\ddag}$     & 0.947$/$0.028$^{\ddag}$    & 22.07$/$3.31$^{\ddag}$     & 0.927$/$0.029$^{\ddag}$    \\
RCGAN                                  & 25.39$/$3.26$^{\ddag}$ & 0.950$/$0.025$^{\ddag}$ & 24.52$/$4.90$^{\ddag}$     & 0.901$/$0.060$^{\ddag}$    & 23.37$/$4.55$^{\ddag}$     & 0.880$/$0.139$^{\ddag}$    \\
IRNet                                        & \textcolor{red}{28.80$/$6.69}$^{\ddag}$ & 0.979$/$0.021 & 30.95$/$4.65$^{\ddag}$     & 0.975$/$0.023$^{\ddag}$    & 29.82$/$2.80$^{\ddag}$     & \textcolor{red}{0.977$/$0.011}$^{\ddag}$    \\
\hline
S2DNets$^{*}$                                     & 25.87$/$5.56$^{\ddag}$ & 0.968$/$0.020$^{\ddag}$ & 27.02$/$3.63$^{\ddag}$     & 0.962$/$0.032$^{\ddag}$    & 26.42$/$2.66$^{\ddag}$     & 0.951$/$0.030$^{\ddag}$    \\
S2DNets                                      & 28.40$/$4.98 & \textcolor{red}{0.979$/$0.015} & \textcolor{red}{31.78$/$4.60}     & \textcolor{red}{0.979$/$0.024}    & \textcolor{red}{30.98$/$3.49}     & 0.969$/$0.024  \\ 
\hline
\end{tabular}
\begin{tablenotes}
       \item $\ddag: p-value<0.01; \dag: p-value<0.05; *:$ S2DNets without TV loss
\end{tablenotes}
\end{threeparttable}
\end{table}
\subsubsection{3.3 Parameter Setting}
\begin{figure}
\includegraphics[width=12.4cm]{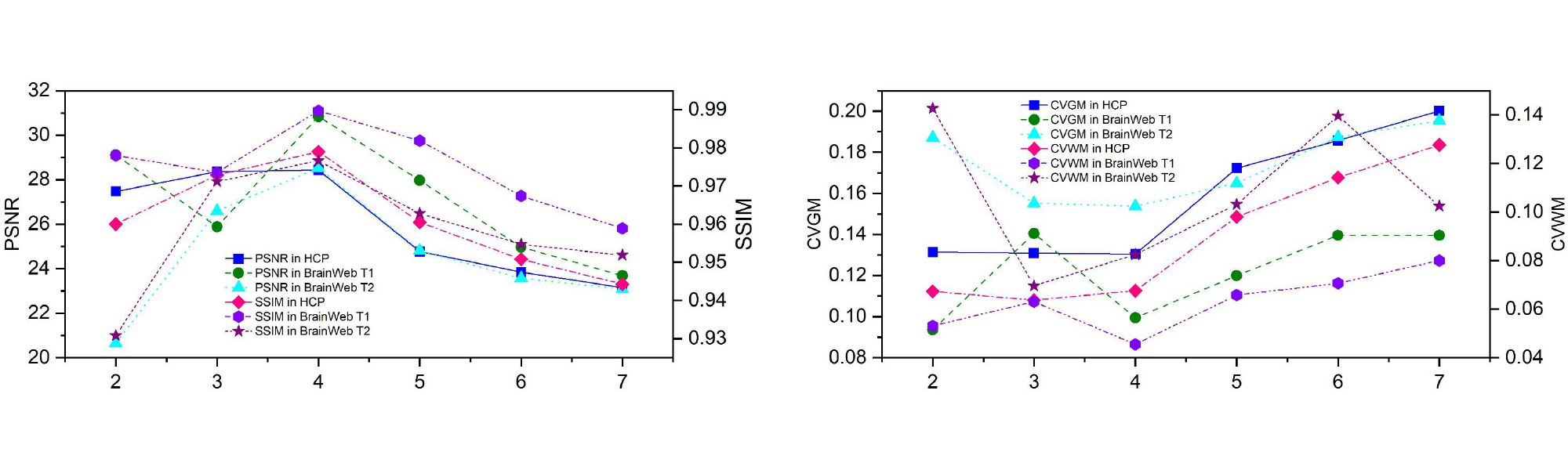}
\centering
\caption{The performance of proposed model on HCP, BrainWeb T1, T2 datasets with different $N$ values.} \label{fig2}
\end{figure}

Fuzziness factor $p$ is generally set to 2 as~\cite{r5}, which is suitable for most of clustering tasks~\cite{r26}. Duan et al.~\cite{r8add} also modeled the local smoothness of bias field by Gaussian Kernel and pointed out that kernel size in $17\times 17$ can deal well with both severe and light intensity inhomogeneity. Besides, they provided the relationship between kernel size $d$ and $sigma$ as: $d\leq 4\times \sigma +1$. Thus, in our experiments, fuzziness factor $p$, kernel size $d$ and shape factor $\sigma$ are set to 2, 17 and 4, respectively. Cluster number $N$ is much more important to influence corrected results. We initialize $N=2$, and gradually increase it by 1 to find appropriate value 4 as Fig~\ref{fig2} shown.
\subsubsection{3.4 Experimental results}
A well-established bias field correction method must meet two essential criteria: 1) the corrected image should maintain sufficient structural details, while ensuring that the intensity distribution aligns with corresponding clean image, as measured by Structural Similarity (SSIM) and Peak Signal Noise Ratio (PSNR); 2) the non-uniform intensity within tissues or organs should be effectively corrected, which can be evaluated by Coefficient of Variation (CV). Table~\ref{tab1} shows the SSIM and PSNR comparison, while Table~\ref{tab2} presents the CV value for white matter (WM) and grey matter (GM). 
\begin{table}[]
\caption{Coefficient of variation (mean/standard deviation) with Wilcoxon-test between proposed and other methods on both \textbf{HCP} and \textbf{BrainWeb} dataset.}
\label{tab2}
\begin{tabular}{lcccccc}\\
\hline\hline
\multicolumn{1}{l}{\multirow{2}{*}{Methods}} & \multicolumn{2}{c}{HCP}    & \multicolumn{2}{c}{BrainWeb T1} & \multicolumn{2}{c}{BrainWeb T2} \\
\cline{2-7}
\multicolumn{1}{c}{}                         & CV$_{GM}$$\downarrow$        & CV$_{WM}$$\downarrow$         & CV$_{GM}$$\downarrow$           & CV$_{WM}$$\downarrow$           & CV$_{GM}$$\downarrow$           & CV$_{WM}$$\downarrow$           \\
\cline{1-7}
Original                                     & 0.254$/$0.068$^{\ddag}$ & 0.194$/$0.080$^{\ddag}$  & 0.240$/$0.062$^{\ddag}$    & 0.210$/$0.074$^{\ddag}$    & 0.283$/$0.054$^{\ddag}$    & 0.224$/$0.073$^{\ddag}$    \\
MICO                                         & 0.136$/$0.044$^{\ddag}$ & 0.075$/$0.026$^{\ddag}$  & 0.201$/$0.056$^{\ddag}$    & 0.162$/$0.062$^{\ddag}$    & 0.244$/$0.046$^{\ddag}$    & 0.186$/$0.068$^{\ddag}$    \\
N4                                           & 0.138$/$0.057$^{\ddag}$ & 0.077$/$0.026$^{\ddag}$  & 0.169$/$0.054$^{\ddag}$    & 0.138$/$0.061$^{\ddag}$    & 0.220$/$0.042$^{\ddag}$    & 0.148$/$0.062$^{\ddag}$    \\
Implicit                                     & \textcolor{red}{0.094$/$0.063}$^{\ddag}$ & 0.062$/$0.052$^{\ddag}$ & 0.130$/$0.035$^{\ddag}$    & 0.089$/$0.023$^{\ddag}$    & \textcolor{red}{0.121$/$0.007}$^{\ddag}$    & 0.098$/$0.046$^{\ddag}$    \\
Yang et al.                                  & 0.137$/$0.048$^{\ddag}$ & 0.094$/$0.048$^{\ddag}$  & 0.186$/$0.049$^{\ddag}$    & 0.158$/$0.057$^{\ddag}$    & 0.237$/$0.041$^{\ddag}$    & 0.174$/$0.058$^{\ddag}$    \\
RCGAN                                        & 0.224$/$0.527$^{\ddag}$ & 0.127$/$0.219$^{\ddag}$  & 0.148$/$0.031$^{\ddag}$    & 0.091$/$0.036$^{\dag}$    & 0.261$/$0.045$^{\ddag}$    & 0.150$/$0.055$^{\ddag}$    \\
IRNet                                        & 0.155$/$0.079$^{\ddag}$ & 0.093$/$0.061$^{\ddag}$  & 0.147$/$0.054$^{\ddag}$    & 0.117$/$0.069$^{\ddag}$    & 0.198$/$0.024$^{\ddag}$    & 0.100$/$0.034$^{\ddag}$    \\
\hline
S2DNets$^{*}$                                       & 0.129$/$0.056$^{\ddag}$ & \textcolor{red}{0.058$/$0.047}$^{\ddag}$  & \textcolor{red}{0.113$/$0.055}$^{\ddag}$    & \textcolor{red}{0.072$/$0.017}$^{\ddag}$    & 0.181$/$0.022   & \textcolor{red}{0.080$/$0.025}$^{\ddag}$    \\
S2DNets                                      & 0.131$/$0.057 & 0.063$/$0.048  & 0.120$/$0.057    & 0.083$/$0.023    & 0.180$/$0.018    & 0.090$/$0.025   \\
\hline
\end{tabular}
\begin{tablenotes}
       \item $\ddag: p-value<0.01; \dag: p-value<0.05; *:$ S2DNets without TV loss
\end{tablenotes}
\end{table}

\begin{figure}[h]
\includegraphics[width=11.6cm]{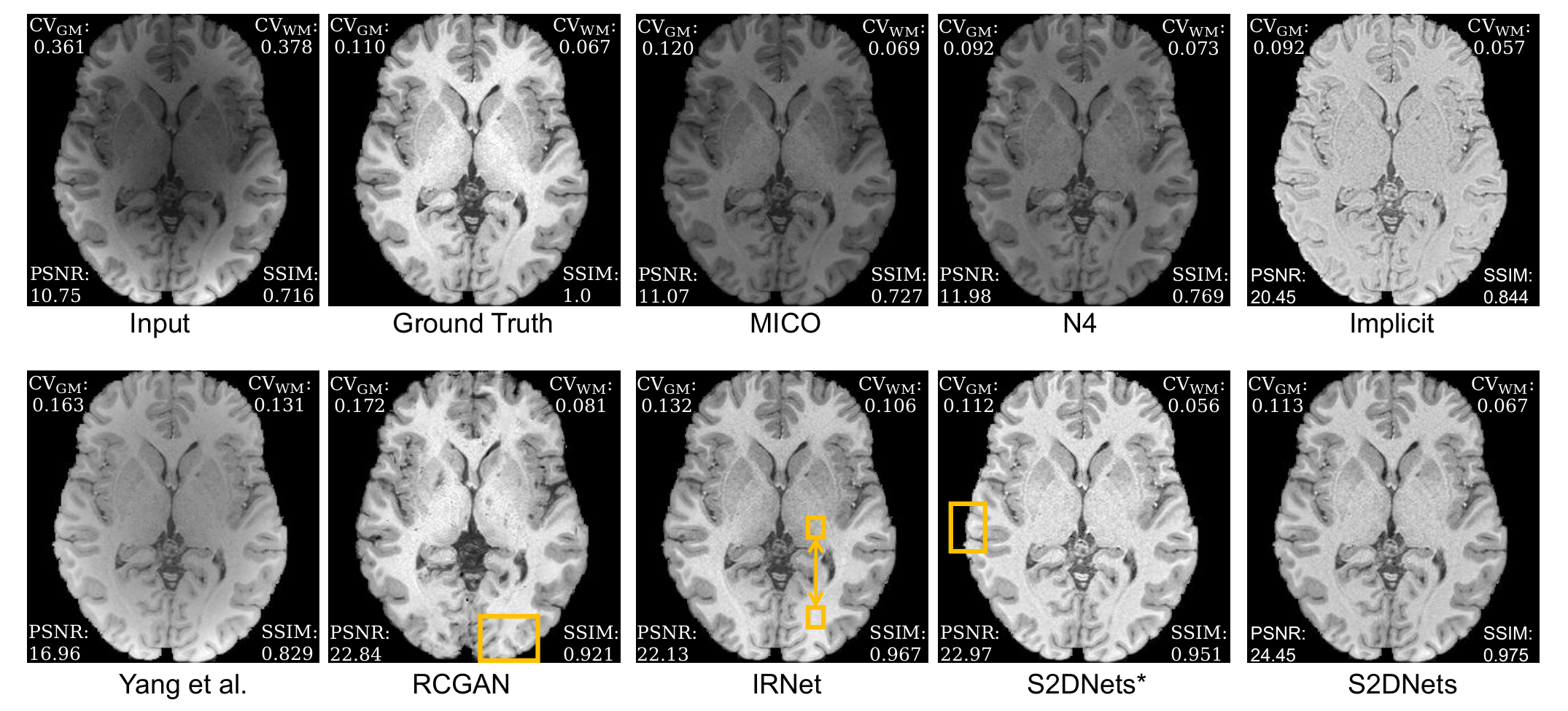}
\centering
\caption{The performance of different methods. Yellow rectangle denotes remained bias field or distorted structure.} \label{fig3}
\end{figure}

From Table~\ref{tab1} and Table~\ref{tab2}, it can be seen that: MICO method effectively removes bias fields with simple intensity distributions in the HCP dataset but struggles with complex bias fields in the BrainWeb dataset. N4 method mitigates bias field by maximizing high-frequency information, but neglects intensity feature, leading to intensity-shift results with lower PSNR and SSIM scores. The Implicit network achieves lower CV values but suffers from over-correction. Yang et al's model improves brightness of MR image but fails to address bias field removal. RCGAN and IRNet both utilize unpaired clean data for supervision. However, due to a lack of smoothness constraints, RCGAN distorts image details, while IRNet, although improving PSNR (0.4 higher in HCP) and SSIM (0.008 higher in BrainWeb T2) fails to introduce the structural constraints of corrected image and misidentifies the bias field as part of the image structure, yielding sub-optimal results with higher CV values. These trends are further confirmed in Fig~\ref{fig3}.

\begin{table}[]
\caption{Dice (\%, mean/standard deviation) segmented by pre-trained 2D U-Net model and software FSL with Wilcoxon-test between proposed and other top-4 methods on \textbf{HCP} dataset.}
\label{tab3}
\begin{tabular}{llcccccc}\\
\hline\hline
\multicolumn{2}{l}{Method}                                                             & Original    & MICO        & N4          & IRNet       & S2DNets$^{*}$     & S2DNets     \\
\hline
\multirow{2}{*}{\begin{tabular}[c]{@{}c@{}}U-Net\end{tabular}} & GM & 63.71$/$13.24$^{\ddag}$ & 83.66$/$12.51$^{\dag}$ & 83.29$/$9.95$^{\ddag}$  & 82.82$/$14.25$^{\ddag}$ & 82.39$/$10.17$^{\ddag}$ & \textcolor{red}{84.16$/$10.31} \\
                                                                                & WM & 49.69$/$18.69$^{\ddag}$ & 83.87$/$15.90$^{\ddag}$ & 83.36$/$13.60$^{\ddag}$ & 79.66$/$21.14$^{\ddag}$ & 83.17$/$14.01$^{\ddag}$ & \textcolor{red}{85.03$/$13.36} \\
                                                                                \hline
\multirow{2}{*}{FSL}                                                            & GM & 47.51$/$12.93$^{\ddag}$ & 69.06$/$21.15$^{\ddag}$ & 67.18$/$16.27$^{\ddag}$ & 69.60$/$22.52      $^{\ddag}$ & 65.14$/$15.38$^{\ddag}$ & \textcolor{red}{70.67$/$15.03} \\
                                                                                & WM & 63.81$/$16.22$^{\ddag}$ & 78.94$/$15.49$^{\ddag}$ & 74.80$/$13.65$^{\ddag}$ & 79.11$/$14.51       & 77.12$/$14.13$^{\ddag}$ & \textcolor{red}{79.11$/$11.58}\\
\hline
\end{tabular}
\begin{tablenotes}
       \item $\ddag: p-value<0.01; \dag: p-value<0.05; *:$ S2DNets without TV loss
\end{tablenotes}
\end{table}

In contrast, S2DNets effectively remove bias fields while preserving structural details with high fidelity. The ablation study on $loss_{TV}$ reveals that while CV values increase slightly (no more than 0.011), PSNR improves significantly by $9.8\%$, $17.6\%$ and $17.26\%$ on HCP, Brainweb T1 and T2 datasets, respectively. This highlights the effectiveness of global smoothness provided by $loss_{TV}$. Finally, the non-parametric Wilcoxon-test indicates significant differences between S2DNets and other methods, with most $p-value$s smaller than predefined threshold (0.01). 
We also evaluate our method on downstream segmentation tasks. We applied a 2D U-Net model, which was pre-trained on clean MR images, and a software FSL~\cite{r28} for brain image segmentation on clinical dataset HCP by Dice metric. As shown in Table~\ref{tab3}, our method obtains the best segmentation performance compared with other top-4 methods.

\section{Conclusion}
We introduce both structure and smoothness constraints into bias field correction framework in a self-supervised learning scheme. The experiments executed on both clinical and simulate dataset show that the proposed method achieves an accurate bias field estimation for MR image recovery and has a wide potential application in the future.


\end{document}